\documentstyle[preprint,eqsecnum,aps]{revtex}
\tighten
\begin{document}

\title{Force-Free Interactions and Nondispersive  \\ 
Phase Shifts in Interferometry} \author{Murray Peshkin 
\footnote{Physics Division, Argonne National Laboratory, Argonne,
IL 60439-4843 USA}}

\maketitle
\bigskip


\begin{abstract}
Zeilinger's observation that phenomena of the Aharonov-Bohm type  
lead to nondispersive, i.e. energy-independent, phase shifts in 
interferometers is generalized in a new proof which shows that the 
precise condition for nondispersivity is a force-free interaction.  
The converse theorem is disproved by a conceptual counter example.  
Applications to several nondispersive interference phenomena are reviewed 
briefly.  Those fall into two classes which are objectively distinct 
from each other in that in the first class phase shifts depend only on 
the topology of the interfering beam paths, while in the second class 
force-free physical interactions take place at identifiable points 
along the path.  Apparent disagreements in the literature about the 
topological nature of the phenomena in the second class stem from 
differing definitions.
\end{abstract}

\section{Introduction}

A. Zeilinger \cite{zeilinger86} 
observed that a nondispersive, i.e.\ energy 
independent, phase shift provides an operational signature of what 
he called Aharonov-Bohm effects in neutron interferometry.  In his 
words,

\begin{quote}
''...the significant operational feature of the AB effect is that 
there exists no observable defined on either beam separately which 
is influenced by the electric or magnetic field.  This is related to 
the property that the AB phase shifts are nondispersive.  Only a 
constant overall phase change for the wave packet arises...''
\end{quote}

That result has proved to be useful to experimenters in confirming 
the results of theories involving effects of the AB type and it has 
also been used as a test in characterizing such effects.

Zeilinger proved his nondispersivity theorem by listing the 
calculated phase shifts in each of what he called the generalized 
Aharonov-Bohm experiments with neutrons and observing that they were 
all nondispersive.  

My purpose here is to expand on Zeilinger's observation in several 
ways, to give a brief review of its application to interferometry 
measurements on neutrons and other particles, and to clarify the 
relations between nondispersivity, force-free motion, and what have 
come to be called topological effects in quantum mechanics.  Section II
presents a new proof from the Schroedinger equation that force-free 
interactions imply nondispersive phase shifts.  This new proof is more 
general than the original one in that it allows for a wide class of 
nonlocal interactions, and it demonstrates that an objective sufficient 
condition for nondispersivity is that the motion of the particles 
should be force free.  In Section III, I show by counter example that 
the converse theorem is not true; a nondispersive phase shift does not 
imply force-free motion in the absence of other assumptions about the 
interactions.  Section IV contains a brief review of the application 
of the nondispersivity theorem to idealized versions of various kinds 
of interferometry experiments.  In Section V, I separate the force-free 
interactions into two classes according to the presence or absence of 
Maxwell fields or other scattering interactions acting on the 
particles and I try to clarify the relation between nondispersive 
phase shifts, force-free motion, and topological interactions.

\section{Zeilinger's nondispersivity theorem}   

Suppose that charged or neutral particles, in a single spin state 
for simplicity, traverse a Mach-Zehnder interferometer (Fig.\ 1) 
and the relative phase of the two partial waves is measured by 
observing the intensities of the emergent beams, O and H.  For 
present purposes, it is adequate to describe the motion of the 
particles in either arm through a wave function in one dimension, 
subject to the Hamiltonian

\begin{equation}
H = {p^2\over 2m} + V \,.
\end{equation}
Here $V$ represents a general interaction, possibly time dependent 
and possibly momentum dependent or otherwise nonlocal, but confined 
at all times to some interaction zone $0\,\small{\leq} {\it x} \small{\leq}\,\ell$.   

The interaction is taken to be entirely elastic; no particle exchanges 
energy with the applied field represented by $V$.  For convenience, 
I postulate temporarily that no particles are reflected from the 
interaction zone.  That restriction is justified below.  Then plane-wave 
particles incident from the left with momentum $p = \hbar k$ will emerge on the 
right in a plane wave state having the same amplitude and the same 
momentum $p$, but possibly shifted in phase by some amount $\delta (k)$.

\begin{eqnarray}
\phi _k (x,t) =
\left\{
\begin{array}{l}
exp\Big\{ i(kx - \omega t)\Big\} \,\,\,\,\,\,\,\,\,\,\,\,\,\,\,\,\,\,\,\,\,\,\,
{\rm for} \,\ x< 0 \\
exp\Big\{ i(kx - \omega t + \delta (k)\Big\} \,\,\,\,\,\,\, {\rm for}\,\ x>\ell\,,
\end{array}\right.
\end{eqnarray}
where $\omega =\hbar k^2/2m$.

A wave packet incident from the left at time zero,

\begin{equation}
\psi (x,0) = \int dk \chi (k) exp\{ikx\}
\end{equation}
will emerge at a later time $T$ when it has completely cleared the
interaction zone as

\begin{equation}
\psi (x,T) = \int dk \chi (k) exp \Big\{ i(kx - \omega T) + i\delta (k) \Big\} \,.
\end{equation}
The expectation of $x$ at time $T$ is given by

\begin{eqnarray}
\langle x\rangle _T &=& \int\int\int dx\, dk\, dk' \chi (k')^*
x\chi (k) exp\Big\{ -i(k' x - \omega ' T) - i\delta (k')\Big\}\\
\nonumber
&\times& exp\Big\{ i(kx - \omega \,T) + i\delta (k) \Big\} \\
\nonumber
&=& \int dk \chi (k)^* exp\Big\{ i\,\omega \,T - i\delta (k) \Big\}
\Bigg( i{\partial \over \partial k} \Bigg) \bigg[ \chi (k)
exp \Big\{ ( -i\,\omega \,T ) + i\delta (k) \Big\} \bigg]
\end{eqnarray}
$\langle x \rangle _0$ is the same, but with 0 substituted for $T$ 
and for the phase shift $\delta (k)$.

\begin{eqnarray}
\langle x \rangle _T &=& i \int \chi (k)^* {d\chi \over dk} \, dk + T \int
\Big| \chi (k) \Big| ^2 {d\omega \over dk} \, dk + \int \Big| \chi (k) \Big| ^2 \,
{d\delta \over dk} \,dk \\
\nonumber
&=& \langle x \rangle _0 + \langle v \rangle _0 \,T + \int
\Big| \chi (k) \Big| ^2 {d\delta \over dk} \, dk \,,
\end{eqnarray}
where $\langle v \rangle$, the expectation of the group velocity, 
is given by

\begin{equation}
\langle v \rangle = {d \over dt} \, \langle x \rangle =
\Bigg\langle {d\omega \over dk} \Bigg\rangle \,.
\end{equation}

Ehrenfest's theorem tells us that if the interaction is force free, then

\begin{equation}
\langle x \rangle _T = \langle x \rangle _0 + \langle v \rangle _0 \, T \,.
\end{equation}
Since the last term in Eq.\ (2.6) must then vanish for all wave packets
$\chi (k)$, it follows that

\begin{equation}
{d\delta \over dk} = 0
\end{equation}
for all k.  That result is Zeilinger's nondispersivity theorem.  
The proof given here assumed no reflected wave, but that condition 
in fact follows from Ehrenfest's theorem in the absence of any force 
on the particle; the expectation of the momentum  cannot be conserved 
in the presence of a reflected wave unless the transmitted wave has 
greater kinetic energy than the incident wave, contrary to the 
assumption of no energy transfer.

Ehrenfest's theorem does not require that the force $F(x,t)$ obeys $F=0$
for all $x$ and $t$ as an operator equation, only that

\begin{equation}
F(x,t) \psi (x,t) = 0
\end{equation} 
for the wave packets in the interferometer.  Therefore, the same 
is true of the nondispersivity theorem.  The particle may traverse 
regions where, for example, a potential gradient existed at earlier 
times, as long as the potential is spatially uniform while the particle 
is present.\footnote{In fact, the particle must traverse such a region 
to acquire a phase shift in the absence any force on the particle
\cite{peshkin89}.}

\section{The Converse Theorem} 
The converse theorem, that a nondispersive phase shift at all energies 
implies a force-free interaction, does not follow from the Schroedinger 
equation, and it is in fact false.  For a simple example to show that 
nonvanishing forces can cause a nondispersive phase shift, consider a 
nuclear phase shifter of the kind commonly used in neutron interferometry.  
A slab of some material is placed in one beam to alter its phase relative 
to the other beam.  The material in the phase shifter is characterized by 
an energy-dependent index of refraction $\eta (k)$ or equivalently by an 
energy-dependent interaction term in the Hamiltonian given by

\begin{equation}
V = {\hbar ^2 k ^2 \over 2m} \, \Big( 1 - \eta \,(k) ^2 \Big)
\end{equation}
for values of $x$ within the slab, and by $V = 0$ elsewhere. 
For this illustration I suppose that $V$ is positive and $\eta < 1$ so 
that no bound states will be introduced.  

The phase shift in a slab of thickness $b$, 

\begin{equation}
\delta = kb \Big( \eta \,(k) - 1 \Big)
\end{equation}
can in principle be made nondispersive by using a material with 
index of refraction given by

\begin{equation}
n(k) = 1 + \delta \Big/ kb
\end{equation}
with constant negative $\delta$.  However, the neutron does experience forces 
when it enters and exits the material.

A uniform slab whose index of refraction is given by Eq.\ (3.3) is 
far from being the only example of a nondispersive interaction 
that does exert forces, but physically they are all rather contrived.

\section{Applications of the Nondispersivity Theorem}
The nondispersivity theorem has been applied to five distinct kinds of 
interferometry experiments performed or proposed:

\subsection{The Magnetic Aharonov-Bohm Effect}

In an idealized interferometric version of the magnetic Aharonov-Bohm 
(AB) effect \cite{aharonov59}
the current in a solenoid placed between 
the two arms 
of the interferometer in Fig.\ 1 is the source of a vector potential $A$.  
Electrons in the interferometer are exposed to $A$  but they are subjected 
to no force because they are confined to regions where the magnetic field 
$B$ vanishes.  The nondispersivity theorem can be applied directly to that 
case by choosing the gauge so that $A$ vanishes outside of the 
interaction zone.  Then

\begin{equation}
V = {e \over mc} \, p \cdot A
\end{equation}
and the phase shift is given by

\begin{equation}
\delta = \int V\, dt = {e \over \hbar c} \int A \cdot dx = {e\Phi \over \hbar c}\,,
\end{equation}
$\Phi$ being the magnetic flux through the solenoid.  Experiments equivalent 
to this have been reported 
\cite{chambers60,bayh62,mollenstedt62,peshkinII}.
The most precise \cite{peshkinII} 
agreed with the calculated phase shifts to better than one percent.

\subsection{The Electric Aharonov-Bohm Effect} 
The electric version of Aharonov-Bohm effect \cite{aharonov59}
is in principle a second direct example, although the practical 
obstacles to carrying out such an experiment are formidable.  
There, each arm of an electron interferometer is surrounded in the 
interaction zone by a cylindrical conducting shield.  While the partial 
wave packets in both arms are inside their respective conducting cylinders, 
the two cylinders are caused to have a potential difference $\Delta \phi (t)$.  
That leads to a phase difference

\begin{equation}
\delta = \int V\, dt = {e \over \hbar} \int \Delta \phi (t) dt\,,
\end{equation}
again independently of the energy of the electrons.

\subsection{The Aharonov-Casher Effect} 
In an idealized interferometric version of the Aharonov-Casher (AC) 
effect \cite{aharonov84,anandan89}
atoms or neutrons with magnetic moment $\mu$ 
move in the $x$ direction with their spins polarized in the $z$ direction.  
In the interaction zone, they traverse an electric field of strength $\pm E$   
in the $z$ direction, $+$ in one arm of the interferometer and $-$ in the other.

The interaction term in the Hamiltonian is given for nonrelativistic
motion by

\begin{equation}
V = {\mu \over mc} \,\sigma \cdot p \times E = \pm \,{\mu E \over mc} \, p\,.
\end{equation}
That leads to force-free motion with phase shifts given by

\begin{equation}
\delta _{\pm} = \int V _{\pm} dt = \pm \,{\mu E\over c} \int {p \over m} \, dt
 = \pm {\mu E\ell \over c} \,.
\end{equation}
The relative phase shift between the two arms is 

\begin{equation}
\delta = 2 E\ell \, {\mu \over c} \,,
\end{equation}
independently of the energy.  This result has been confirmed in neutron 
interferometry experiments \cite{cimmino89,kaiser91} 
and with high precision 
in atom interferometry experiments \cite{sangster93,sangster95} 
that demonstrated the nondispersivity over a wide range of energies.

\subsection{The Scalar Aharonov-Bohm Effect}
In what has been called the scalar Aharonov-Bohm (SAB) effect, 
neutrons in the interaction zone traverse an external magnetic field $B(t)$.  
In the simplest case, the neutrons are polarized in the direction of $B$.  
Ideally, $B(t)$ should be turned on and then off while the neutronÕs wave 
packet is inside the interaction zone, to avoid forces due to grad $B$ at the 
edges of the field.  The phase shift is then given by

\begin{equation}
\delta = \int V\,dt = - \,{\mu \over \hbar} \int B(t)\,dt \,,
\end{equation} 
independently of the energy of the neutron.  Experiments doing the practical 
equivalent of that have been reported, both with unpolarized \cite{allman92} 
and polarized \cite{badurek93,lee98} 
neutrons, and did find the expected 
phase shifts.  In neutron interferometry, the range of neutron energies 
is narrow, but the nondispersivity can nevertheless be useful 
because it guarantees that the phase shift is uniform across the wave packet.

\subsection{The Force-Free Nuclear Phase Shifter}
Conceptually, it is possible to create a phase shifter for neutron 
interferometry by having a gas cell in the interaction zone.  
Some gas is pumped into the cell and then pumped out again, all while 
the neutron's wave packet is entirely within the interaction zone. 
This results in a time-dependent index of refraction, or equivalently a 
time-dependent force-free interaction potential $V$, and therefore a 
nondispersive phase shift analogous to the ones in the AC and SAB 
effects and sharing their Schroedinger equations under appropriate 
substitution in the interaction term $V$.  The force-free nuclear phase 
shifter was introduced as a thought experiment \cite{peshkin95}
to separate 
the question of force-free motion from issues involving electromagnetism, 
gauge fields, and torques on spinning particles.

\section{Conclusions}
Nondispersive phase shifts are a signature of force-free motion.  
A nondispersive phase shift at all energies is required for, but in
principle does not imply, force-free motion.  However, for motion 
under the influence of forces to exhibit a nondispersive phase shift 
at all energies seems improbable in a real case.

	There are two objectively different classes of force-free interactions in 
interferometry, both nondispersive.  In the first class are the magnetic and 
electric Aharonov-Bohm effects.  There, the interaction is with a gauge 
field and a gauge transformation can move the location of the interaction 
from one arm of the interferometer to another.  

	The second class contains the AC and SAB effects and the force-free 
nuclear phase shifter.  There the physics is significantly 
different \cite{peshkin95,goldhaber89}.  
The interaction occurs 
in an objectively known place.  In the case of AC and SAB there are torques 
whose expected values vanish but whose quantum fluctuations do not and 
there are observables, measurable in principle, that respond to those torques.  
In the nuclear phase shifter there is an ordinary potential energy.  
	
	A nondispersive phase shift has often been cited as evidence of a 
topological interaction.  That is true if a topological interaction is 
defined to be a force-free one.  A possibly unwanted consequence of that 
definition is that it includes the interaction with the nondispersive 
nuclear phase shifter as a topological interaction, but not the interaction 
with an ordinary nuclear phase shifter with very similar physical properties.

The AB effect is thought to be a topological phenomenon for the reasons 
given above and because it has an important relation to Berry's 
geometrical phase \cite{berry84},
which is topological in nature.  
The literature contains many discussions or statements about the 
topological nature of the interactions in the second class \cite{pfeifer94}.
Their lack of agreement appears to arise entirely from the use of different, 
sometimes unstated, definitions.

This work supported by the U. S. Department of Energy, Nuclear Physics Division,
under contact W-31-109-ENG-38.

\begin{figure}
\caption{Mach-Zehnder interferometer shown with an interaction zone of length
$\ell$ in each arm.  The solid circle represents a solenoid used for 
the magnetic Aharonov-Bohm effect.}
\end{figure}

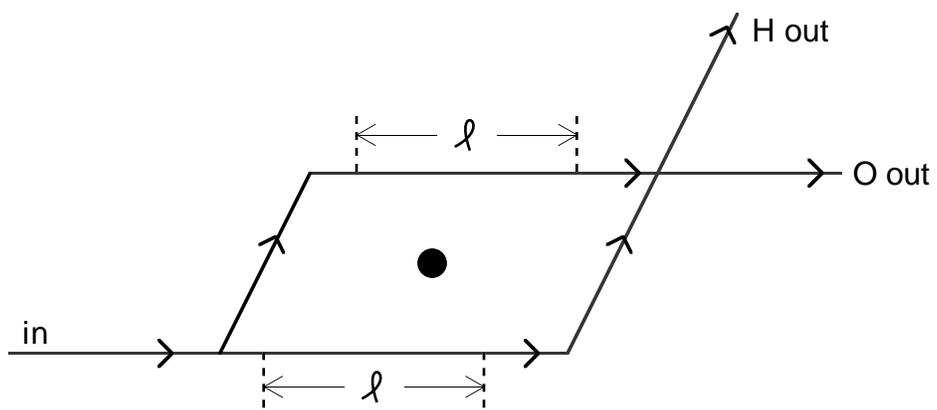
\end{document}